# Thermal and Electrical Properties of Hybrid Composites with Graphene and Boron Nitride Fillers

**Jacob S. Lewis, Zahra Barani, Andres Sanchez Magana, Fariborz Kargar and Alexander A. Balandin***

Phonon Optimized Engineered Materials (POEM) Center, Department of Electrical and Computer Engineering, Materials Science and Engineering Program, Bourns College of Engineering, University of California, Riverside, California 92521 USA

*Corresponding author (A.A.B.): balandin@ece.ucr.edu ; web-site:  http://balandingroup.ucr.edu/

# Abstract

We report on the thermal and electrical properties of hybrid epoxy composites with graphene and boron nitride fillers. The thicknesses, lateral dimensions, and aspect ratios of each filler material were intentionally selected for geometric similarity to one another, in contrast to prior studies that utilized dissimilar filler geometries to achieve a "synergistic" effect. We demonstrate that the electrically-conductive graphene and electrically-insulating boron nitride fillers allow one to effectively engineer the thermal and electrical conductivities of their resulting composites. By varying the constituent fraction of boron nitride to graphene in a composite with ~44% total filler loading, one can tune the thermal conductivity enhancement from a factor of ×15 to ×35 and increase the electrical conductivity by many orders of magnitude. The obtained results are important for the development of next-generation thermal interface materials with controllable electrical properties necessary for applications requiring either electrical grounding or insulation.

# INTRODUCTION

Suitably reliable and affordable thermal management technologies remain a major challenge driven by the continuous miniaturization of integrated circuits, increasing functionality of mobile devices, and growing computing density in data centers [1]. Development of the next generation of thermal interface materials (TIMs) is important for packaging and thermal management of monolithic and stacked integrated circuits, memory devices, microwave sources, light-emitting devices, and solar cells [2–7]. Improved TIMs promise to provide a low-cost solution to help manage hot spots at thermal densities over 500 W/cm$^2$ [1]. Although existing conventional polymeric TIMs facilitate thermal conductance through otherwise primarily air-gapped solid-on-solid junctions, there is great desire for and room for improvement in curing and non-curing TIMs with substantially increased thermal conductivity [8,9]. Existing commercially-available TIMs, also referred to as thermal adhesives or thermal greases, have a bulk thermal conductivity ranging from 0.5 W/mK to 5 W/mK, with values usually around 1.0 W/mK [9]. Achieving high thermal conductivity in TIMs is often accompanied by an increase in electrical conductivity resulting from the use of electrically conductive carbon-based or metallic fillers [10–12]. TIMs are used in various applications that require specific electrical conductivity levels spanning from overall insulation to conduction. Methods that would provide for the direct and detailed control of TIM electrical conductivity are of substantial practical importance.

The unique heat conduction properties of graphene, discovered a decade ago [13–16], stimulated investigations into graphene's potential as a filler material in cured and non-cured TIMs [8,17–21] and solid thermal coatings [22–24]. Initial studies were limited to small loading fractions, $f<10$ vol. %, of graphene fillers yet still achieved thermal conductivities of ~5 W/mK near room temperature (RT) [8,25]. Recently, epoxy composites with high loadings of graphene, $f>40$ vol. %, have been demonstrated by several groups, reaching thermal conductivities of ~12 W/mK [19,25–27]. In high loading graphene composites, both electrical and thermal percolation thresholds have been met, resulting in a marked increase in both thermal and electrical conductivity. One possible approach for enhancing the heat conduction in a composite without inducing electrical conductivity is through mixing graphene fillers with other thermally conductive but electrically insulating fillers. Hexagonal boron nitride ($h$-BN) is a natural choice for use as a complementary, electrically insulating, filler with graphene [26,28–30]. It is known that $h$-BN has a high intrinsic thermal conductivity [31–33]. Because of $h$-BN's long use as a solid lubricant, it benefits from a mature, low-cost industrial production that is essential for any thermal management materials. To date, studies concerned with both graphene and $h$-BN fillers were either limited to low total loading fractions [28,30,34] or utilized extremely dissimilar sizes and aspects ratios of graphene and $h$-BN fillers [26,29]. Here we report on the study of composites with high hybrid loadings of graphene and $h$-BN fillers, which were intentionally kept at similar thickness, lateral dimensions, and aspect ratios.

## EXPERIMENTAL SECTION

### Composite Material Design

We start by establishing the terminology used in this study. Unlike the strict definition of graphene being a monolayer of hexagonally-bound carbon atoms found in the solid-state physics context, the term *graphene* in the practical field of thermal composites refers to a mixture of "true" graphene and few-layer graphene flakes [8,35]. Typical flakes in such graphene mixtures have a thickness of a few nanometers and lateral dimensions of a few micrometers. Acoustic phonons are the main heat carriers in graphene despite its high electrical conductivity [36,37]. As a result, it is essential for heat conduction that flake lateral sizes be above a micrometer because the grey mean free path (MFP) for acoustic phonons in graphene near RT is ~750 nm [14]. As defined, graphene used in TIMs is different from graphene nano-platelets, *i.e.* particles with small lateral dimensions and aspect ratio, or milled graphite, *i.e.* particles with large lateral dimensions and thicknesses in the micrometer range. The same convention in terminology will be used for $h$-BN fillers in the context of epoxy composites. It was previously established that the optimum composition of graphene fillers in terms of distribution of their thicknesses and lateral dimensions depends on a number of fundamental and technological considerations [8,14,19]. For example, single-layer graphene has the highest intrinsic thermal conductivity. However, the thermal conductivity of "true" graphene will experience the strongest degradation upon exposure to the matrix material. Long graphene fillers with dimensions above the average phonon MFP will better preserve the heat conduction properties. However, when the filler length becomes too large it leads to its bending and can start to interfere with the bond line thickness requirements. The intrinsic properties of electrically conductive graphene fillers and electrically insulating $h$-BN fillers, reported in literature [13,14,31,38–47], are summarized in Table I.

[Table I: Intrinsic Properties]

The use of two or more types of fillers in a single composite is a well-established strategy in the design of TIMs (see Figure 1). A combination of two fillers can provide a "synergistic" enhancement of the thermal conductivity, superior to that achievable with any composite composed of either filler alone [48–57]. There are indications that the "synergistic" effect occurs when each filler is dissimilar to the other in size and aspect ratio. This observation can be explained by a tendency of a smaller filler fitting between larger flakes, providing a path between the two large flakes in which heat may flow between the large flakes with no or less distance traversed through the highly insulating matrix. In another scenario, high aspect ratio fillers, $e.g.$ carbon nanotubes, can efficiently connect large spherical metal particles even when the loading of carbon nanotubes is relatively low [48]. For this reason, previous attempts at combining graphene with other fillers were focused on interfacing fillers of distinctively different size and aspect ratio [17,26,29]. In Figure 1 (a-d), we illustrate the conventional approaches for composite materials' design utilizing fillers of different sizes and aspect ratios [48–57]. In the present work, we take an unconventional approach, and use graphene and $h$-BN fillers of approximately the same thicknesses, lateral dimensions, and aspect ratios, as shown in Figure 1 (e-f). The fundamental goal is to investigate if substantial enhancements of thermal conductivity for both electrically conductive and electrically insulating TIMs are possible with similar size fillers, without inducing the "synergistic" effect. The practical goal is to achieve a greater thermal conductivity than that of commercially-available TIMs while a using lower total filler loading. The similarity in the filler dimensions should bring additional benefits of more simple processing and reduced cost.

[Figure 1: Composite Design]

## Material Synthesis

The polymer base, *i.e.* matrix material, is a room-temperature curing epoxy composed of a resin (Bisphenol-A; Allied High Tech Products, Inc.) and a hardening agent (Triethylenetetramine; Allied High Tech Products, Inc.). The graphene flakes were sourced from a commercial liquid-exfoliated graphene powder (Graphene Supermarket) having a thickness ranging from 0.35 nm, corresponding to specimens of a single atomic plane, to 12 nm and lateral sizes ranging from 2 μm to 8 μm. The *h*-BN filler were sourced from a similar process (US Research Nanomaterials, Inc.) with a similar range of thicknesses and lateral sizes ranging from 3 μm to 8 μm. The two filler materials were incorporated into the polymer matrix material without extra processing steps such as additional exfoliation or surface functionalization. We intentionally omitted any functionalization to facilitate a more direct comparison of the effects graphene and *h*-BN fillers have on resulting composites. One should note that functionalization, filler size optimization, and directed orientation lead to further improvements in the overall thermal conductivity of the resulting composites [8,18,19,58–60].

The epoxy composite preparation began with weighing the resin material, pre-calculating, and weighing of requisite filler materials necessary to hit a targeted composite filler loading level. Next, the prescribed amounts of fillers were pre-mixed separate from the resin to achieve homogeneity and then added to the resin. Then, the resin and fillers were mixed together for 2 minutes in a high-speed, bladeless mixer (Flacktek, Inc.) that revolves about a central axis while simultaneously rotating about the sample's own center axis at 2500 RPM. For all samples with greater than 11 vol. % filler loading, a process of mixing and breaking agglomerations was conducted with a custom-designed needle-like tool. These two mixing steps were then repeated four times in total. After, the hardening agent was added and the composite was speed mixed at 3500 RPM for ten seconds, followed by extensive manual needle mixing. Finally, a pressure of approximately 16 bars was applied via a plunger pressed into the rigid mold to break the numerous individual agglomerations and flatten the material. The sample was left to cure for 24 hours. Finally, the composite sample was then removed from the mold, polished to the desired geometry, and baked at 50°C for ten minutes to ensure that any moisture from the polishing process was evaporated and the curing process was complete. A typical final sample geometry is a disk with a diameter of 25.6 mm and a thickness of 2 mm (see Figure 2).

## Materials Characterization

Raman spectroscopy experiments were conducted to verify the composition and quality of the composite samples. The spectra were attained using a standard spectrometer (Renishaw InVia) under 633-nm laser excitation in the backscattering configuration. More details of our Raman measurement procedures can be found elsewhere in the study of other material systems [61]. Figure 2 (a) shows Raman spectrum of a composite with 21.8 vol. % graphene and 21.8 vol. % $h$-BN fillers. The presence of graphene's characteristic $G$ and $2D$ peaks confirm the incorporation of graphene in the composites [61, 62]. The disorder-related $D'$ peak could be attributed to the interaction between the graphene fillers and the epoxy matrix. However, we note that pure epoxy has a Raman peak at the same frequency. Thus, the observed feature at 1610 cm$^{-1}$ could be combined peaks from the two sources. Another disorder-related $D$ peak is expected in such composite materials owing to defects in graphene fillers and the effect of the edges [63–65]. The peak at 1365 cm$^{-1}$ was identified as the phonon $E_{2g}$ mode of $h$-BN [66]. The composite samples were also inspected using scanning electron microscopy (SEM). The SEM (Carl Zeiss AG Gemini) analysis was conducted on a freshly fractured and exposed composite surface to show the simultaneous inclusion and distribution of both graphene and $h$-BN fillers. In Figure 2 (b) we present a micrograph of a representative sample. The pseudo-colors are used for clarity to demonstrate graphene fillers (blue and green flakes), $h$-BN fillers (pink regions in upper-middle of micrograph), and dark blue indicating a microscopic void in the fractured surface. The brighter spots on the fillers, in the center of the micrograph, are likely due to the local electrical charge accumulation on the electrically insulating $h$-BN material. The location of individual $h$-BN fillers in this particular SEM image between graphene flakes on both the right and left side suggests that $h$-BN fillers separate graphene fillers from electrical contact, increasing the local electrical resistivity.

[Figure 2: Samples]

Accurate measurement of a sample's mass is important for the eventual determination of the thermal conductivity using transient methods, such as "laser flash" [67,68]. The densities of the samples were measured using Archimedes' principle with a special electronic scale (Mettler Toledo). In this technique, density is determined by measuring the weight of a sample both in and out of water and calculating the density from the measured difference in weight in each medium (see Supplemental Materials). Figure 3 shows the density for all synthesized samples. One should notice the excellent agreement between the mass densities of samples at an individual total filler loading level. This is due to the similar volumetric mass density of graphene and $h$-BN (see Table I). The deviation from this trend for samples of the highest total filler loading levels, $f$=43.6 vol. %, was attributed to the formation of small voids, *i.e.* air gaps, in the samples of high graphene constituent fraction. The high loading samples have a higher viscosity, which complicates mixing, and may result in the formation of voids resulting in a corresponding reduction in the volumetric mass density, compared to what it would be otherwise.

[Figure 3: Mass Density]

## RESULTS AND DISCISSION

### Thermal Conductivity

The cross-plane thermal diffusivities of the samples were measured using the transient "laser flash" technique (Netzsch, LFA 467). In this technique, a light pulse impinges on one surface of a sample creating a temperature differential in the sample with a corresponding heat wave and measures the infrared intensity on the opposing surface over time. Details of the "laser flash" measurement procedures have been reported elsewhere [69]. The thermal conductivity, $K$, is calculated as $K = \alpha \times \rho \times c_p$, where $\alpha$ is the thermal diffusivity, $\rho$ is the volumetric mass density, and $c_p$ is the heat capacity at constant pressure [67]. The heat capacity is determined using the effective medium approximation (EMA) with the known $c_p$=0.807 J/gK for $h$-BN [70] and graphite's heat capacity of $c_p$=0.72 J/gK for graphene [40]. The specific heat of the base epoxy was measured by differential scanning calorimetry (Netzsch, Polyma 214), revealing a $c_p$ of 1.5 J/gK (see Supplementary Materials). Repeated measurements for each sample produced consistent results with deviations of less than 1%.

Figure 4 shows the thermal conductivity of composites as a function of the constituent fraction of graphene at a total filler loading. The total filler loadings, *i.e.* the load level of both graphene and *h*-BN combined, are shown in the legend. Composites shown with 0 % constituent fraction of graphene have the stated total filler loading level purely composed of *h*-BN. Note the monotonic increase in the thermal conductivity of composites with increasing constituent fraction of graphene for each total loading. The data suggest an absence of the "synergistic" effect for binary fillers of the similar size and aspect ratio. In all instances, the greater intrinsic thermal conductivity of graphene compared to *h*-BN proved to be the dominant factor in determining overall composite thermal conductivity [13,33]. This outcome is sensible considering the similarity of all other filler parameters such as lateral sizes, thicknesses, densities, and mechanical flexibility. The data scatter for composites of 25.5 vol. % total loading is attributed to the proximity of this loading level to the percolation threshold, resulting in some samples unpredictably achieving better percolation networks than others [19,71]. The overall values of thermal conductivity are rather high, and exceeds those of typical commercial TIMs, which utilize total filler loadings as high as 80 vol. % [2,4,9].

[Figure 4: Thermal Conductivity]

**Electrical Conductivity**

In order to measure the cross-plane electrical conductivity, $\sigma = 1/\rho$, of the composites, we painted silver electrodes on the top and bottom bases of the sample cylinders, following the procedure described in literature [72,73]. A multi-meter was connected to the electrodes with a test potential of 9 V applied to perform resistance measurements across the samples with a well-defined electrical pathway. In Figure 5, we present the electrical conductivity of the composites as a function of the constituent fraction of graphene in the total filler loading. The total filler loadings, *i.e.* graphene and *h*-BN, are shown in the legend. The composites with 0 % of graphene, again, are purely filled with *h*-BN. For low total filler loading, *f*=11.4 vol.%, the electrical conductivity increases strongly with increasing graphene constituent fractions. For high total filler loadings, the electrical conductivity shows saturation phenomena for graphene constituent fractions above 25%. Note that the electrical conductivity can be varied by over *eleven orders of magnitude* by altering the graphene constituent fraction in an individual total filler loading level. Interestingly, the composites with high loading of pure *h*-BN fillers, *i.e.* without any graphene, revealed measurable electrical conductivity. The latter was attributed to the fact that *h*-BN, as a wide-bandgap III-V semiconductor material with finite electrical conductivity, which leads to electrical conduction via the development of an electrical percolation network in high-loading samples.

[Figure 5: Electrical Conductivity]

**Discussion and Comparison**

In order to put the present results in the general context of graphene-enhanced TIMs, we summarized thermal conductivities of the composites reported in literature in Table II. Overall, the obtained values of *K*=8 W/mK for composites filled purely with graphene fillers and *K*=3.5 W/mK for composites of *h*-BN fillers compare well with modern TIMs. The present composites perform better than commercial TIMs despite lower total filler loadings, and many TIMs with graphene and *h*-BN reported previously. Higher values were reported by an independent group for graphene and hybrid mixtures of graphene and *h*-BN composites [26,29], which were prepared with more processing steps. The absolute values of the thermal conductivity of the composites can increase further with process optimization and filler functionalization. The noticeable difference in thermal conductivity data obtained in this work from prior studies is the absence of "synergistic effect" [48–57]. The "synergistic effect" refers to the enhancement of thermal conductivity in composites with two different fillers beyond that which is obtained in composites with either single filler type at a comparable loading (one filler loading is equal to the total loading of both fillers in the two-filler systems). The "synergistic effect" was attributed to the combined action of two fillers with different sizes and aspect ratios [34,74–77]. Our results add validity to this proposed mechanism by providing evidence for the *absence* of a synergistic effect with the intentional use of different materials but with similar lateral dimensions, thicknesses, and aspect ratios. The strategy used in this work indicates a promising strategy to use a two-filler system with similar filler geometries for the synthesis of epoxy composites with finely controllable thermal and electrical conductivities. The fact that both types of fillers have the same dimensions can simplify the TIM production and cut the associated costs of manufacture.

[Table II: TIM Comparison]

## Conclusions

We investigated the thermal and electrical properties of hybrid epoxy composites with graphene and boron nitride fillers. The thicknesses, lateral dimensions, and aspect ratios of both types of fillers were specifically chosen for their similarity. We demonstrated that the electrically conductive graphene and electrically insulating boron nitride fillers allow one to effectively engineer the thermal and electrical conductivities of the resulting composites. By varying the constituent fraction of graphene in a composite of ~44% total filler loading one can tune the thermal conductivity enhancement from a factor of ×15 to ×35 and change the electrical conductivity to an even greater extent -- from below $10^{-12}$ S/cm to more than 0.1 S/cm. A thermal conductivity of ~ 8 W/mK was obtained with composites purely filled with graphene and ~6.5 W/mK with an equal mixture of graphene and *h*-BN fillers. These values are above those achievable in modern commercial, cured TIMs with high filler loading. The obtained results are important for the development of next-generation thermal interface materials with controlled electrical properties essential for applications demanding either electrical insulation or electrical grounding.

*Acknowledgements*

This work was supported, in part, by NSF through the Emerging Frontiers of Research Initiative (EFRI) 2-DARE award EFRI-1433395, and by the UC-National Laboratory Collaborative Research and Training Program – University of California Research Initiatives LFR-17-477237. The authors thank Alec Balandin (Riverside STEM Academy) for assistance with the composite preparations.

*Contributions*

A.A.B. conceived the idea of the study, coordinated the project, and contributed to the experimental data analysis; J.S.L. synthesized the composites and conducted thermal measurements; Z.B. and A.S.M. contributed to sample preparation and material characterization; F.K. assisted with the project coordination and contributed the data analysis. A.A.B. led the manuscript preparation. All authors contributed to writing and editing of the manuscript.

## CAPTIONS

**Figure 1:** Illustration of different strategies for the use of binary fillers in thermal composites. (a) Hybrid fillers consisting of spherical particles and quasi-1D fibers, *e.g.* carbon nanotube. The system is characterized by fillers of different materials, sizes, and aspect ratios. (b) Hybrid fillers consisting of spherical geometries of different radii. (c) Hybrid fillers consisting of large-aspect ratio quasi-2D fillers, *e.g.* graphene (blue), and small-aspect ratio nanoparticles of a different material, *e.g. h*-BN (brown). (d) Hybrid fillers consisting of large-aspect ratio quasi-2D fillers, *e.g.* large *h*-BN fillers, and spherical nanoparticles of the same material. This system is characterized by fillers of the same material that differ in size and aspect ratio. (e) Hybrid fillers consisting of large-aspect ratio quasi-2D fillers of different materials, *e.g.* graphene (blue) and *h*-BN (brown). This system is characterized by fillers of the same size and aspect ratio but different material. (f) The same as in (e) but with the larger loading of *h*-BN then graphene fillers. Note that the composite in (e) is expected to be electrically conducting while the composite in (f) is expected to be electrically insulating. The strategies (a) – (d) have been investigated for many constituent fillers, including graphene and *h*-BN. The strategies (e) and (f) are proposed in the present work.

**Figure 2:** (a) Raman spectrum of the epoxy composite with 22% graphene and 22% *h*-BN filler loading by volume. The peaks marked as *G, 2D, D* and *D'* are well-known graphene signatures. The peak labeled as $E_{2g}$ is characteristic of *h*-BN. Inset is an optical image of the epoxy composites with the filler loading varying as (from left to right) 0 % graphene and 44% *h*-BN, 22% graphene and 22% *h*- BN, and 44% graphene and 0% *h*-BN by volume. (c) Pseudo-colored scanning electron microscopy image of a fractured region of a composite with 13% graphene and 13% *h*-BN loading by volume. The pink flakes are *h*-BN fillers while the blue or green flakes are graphene fillers. The scale bar is 1 µm.

**Figure 3:** Mass density of the representative composites. The total loading of graphene and *h*-BN fillers is indicated in the legend. The average for each set is shown with the dashed line. Note that the mass density does not change with the varying fraction of graphene. A deviation for the sample

with the total filler loading of 43.6 % is attributed to the changes in the viscosity characteristics and possible formation of small air bubbles.

**Figure 4:** Thermal conductivity of composites as a function of the filler constituent fraction of graphene in the total filler loading. The total filler loadings, *i.e.* graphene and *h*-BN, are shown in the legend. The composites with 0 % of graphene have pure *h*-BN fillers. Note a monotonic increase in the thermal conductivity of the composites with increasing fraction of graphene for each total loading. The data suggest an absence of the "synergistic" effect for binary fillers of the same size and aspect ratio. Graphene outperforms *h*-BN as the thermal filler owing to its higher intrinsic thermal conductivity. The data scatter for a composite with the 25.5% total loading is attributed to the proximity of the loading level to the percolation threshold.

**Figure 5:** Electrical conductivity of composites as a function of the fraction of graphene in the total filler loading. The total filler loadings, *i.e.* graphene and *h*-BN, are shown in the legend. The composites with 0 % of graphene have pure *h*-BN fillers. For the small total filler loading $f$=11.4 vol.%, the electrical conductivity increases strongly with increasing graphene fraction (red circles). For the high total filler loading, the electrical conductivity shows a saturation type behavior for the graphene fraction above 25%. Note that the electrical conductivity can be varied over more than eleven orders of magnitude by changing the graphene fraction in the total filler loading.

**TABLES**

| Table I: Intrinsic Properties of Filler Materials | | | | | |
|---|---|---|---|---|---|
| Material | K (W/mK) | $\sigma$ (S/cm) | $\rho$ (g/cm$^3$) | $C_p$ (J/gK) | Refs. |
| Graphene | $2000 - 5300$ | $10^3$ - $10^6$ | 2.27 | 0.72 | [13,14,41,42,45,46,78,79] |
| $h$-BN | $250 - 400$ | $10^{-11}$ | 2.29 | 0.81 | [31–33,43,47] |
| Epoxy | 0.22 | $10^{-13} - 10^{-15}$ | 1.17 | 1.51 | [38,39] |

| Table II: Thermal Conductivity of Composites with Graphene and *h*-BN | | | | | |
|---|---|---|---|---|---|
| Filler | *f* (vol. % / wt. %) | Matrix | K (W/mK) | Comments | Refs. |
| Graphene / *h*-BN | vol: 21.8 / 21.8 | Epoxy | 6.5 | | This work |
| | vol: 16.0 / 1.0 | Epoxy | 4.7 | | [26] |
| | wt: 6.8 / 1.6 | Polyamide | 0.9 | | [30] |
| | wt: 20 / 1.5 | Polystyrene | 0.7 | | [34] |
| Graphene | vol: 43.6 | Epoxy | 8.2 | | This work |
| | vol: 10.0 | Epoxy | 5.1 | | [8] |
| | vol: 43.6 | Epoxy | 11.0 | | [19] |
| | wt: 10.1 | Polystyrene | 4.0 | | [25] |
| | vol: 24 | Epoxy | 12.4 | | [26] |
| | wt: 20 | Polystyrene | 0.5 | | [34] |
| | wt: 2 | | 0.5 | Functionalized fillers | [80] |
| | wt: 10 | Epoxy | 1.5 | Functionalized fillers | [81] |
| *h*-BN | vol: 43.6 | Epoxy | 3.46 | | This work |
| | vol: 45 | Epoxy | 5.5 | | [19] |
| | wt: 30 | Epoxy | 0.5 | | [82] |
| | vol: 15 | Epoxy | 6.1 | Oriented fillers | [83] |
| | vol: 30 | Silicone | 5.5 | Oriented fillers | [84] |
| | vol: 9 | Silicone | 0.6 | Oriented fillers | [85] |
| | vol: 44 | Epoxy | 9.0 | Oriented fillers | [86] |

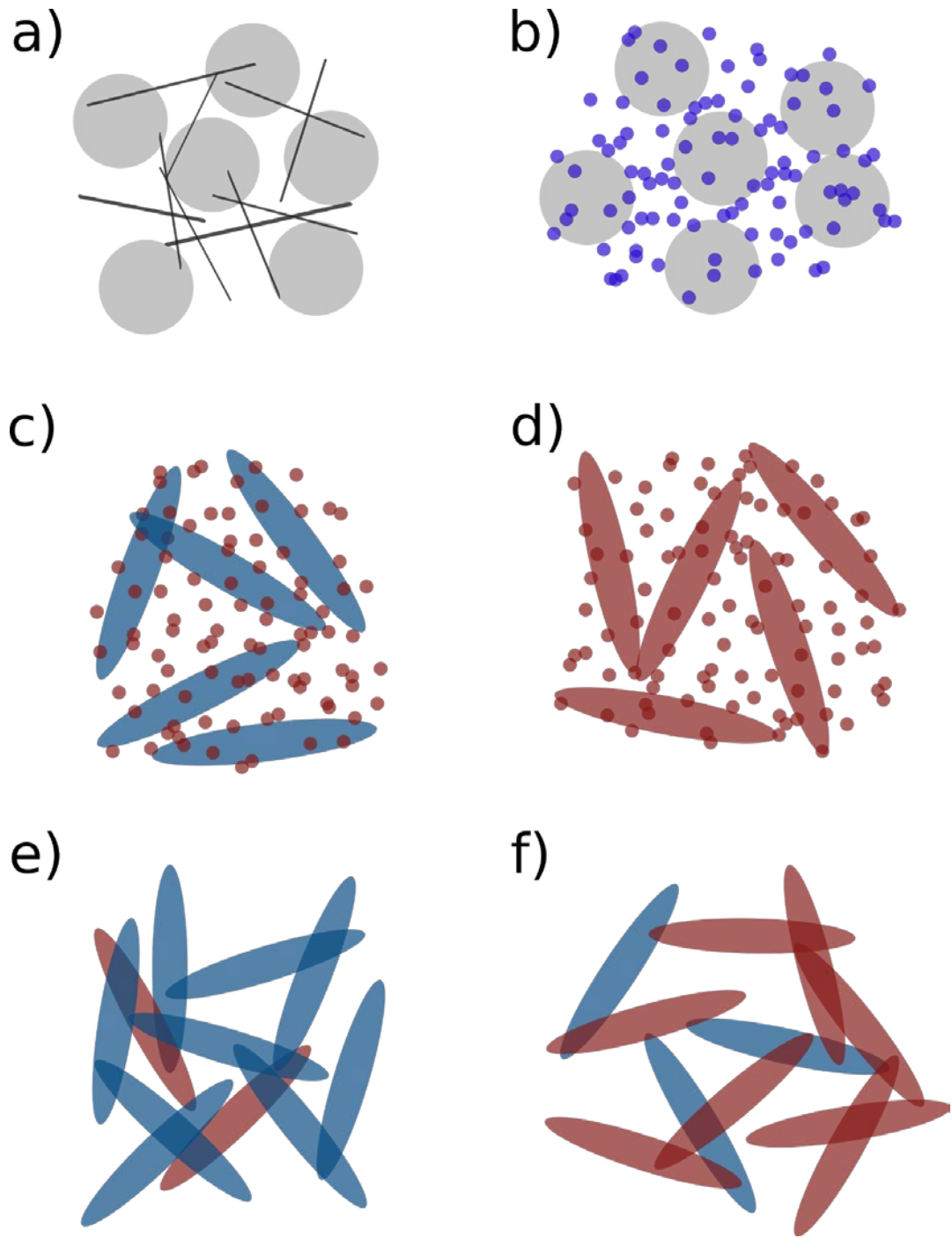

FIGURE 1

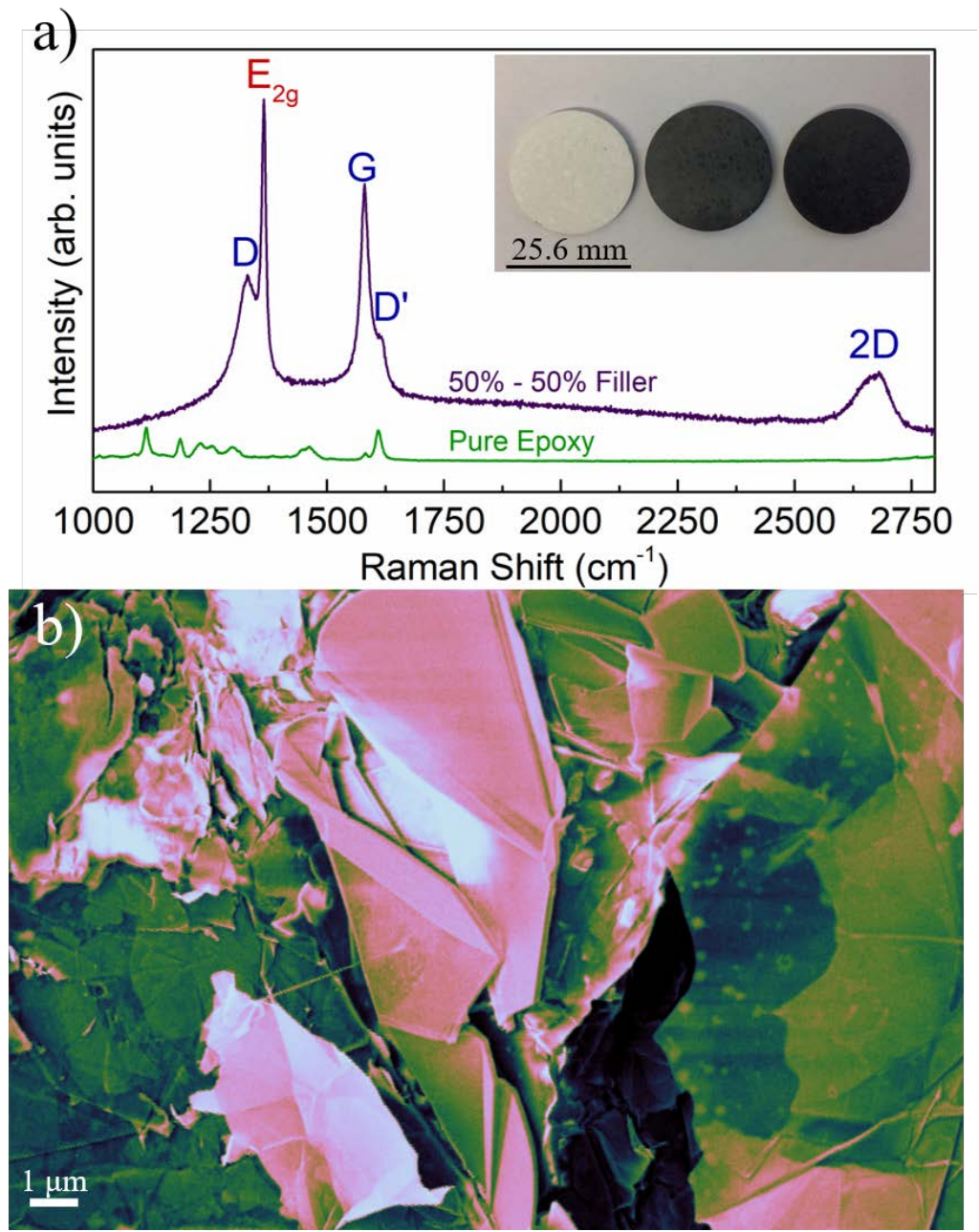

FIGURE 2

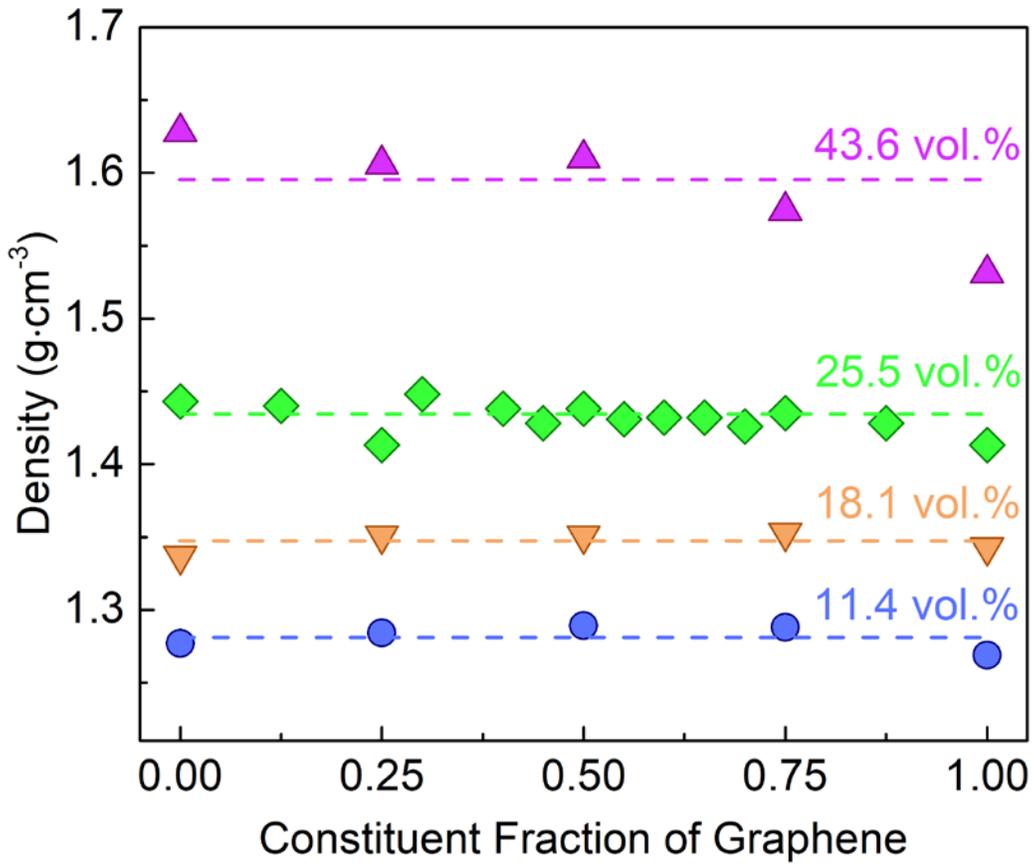

FIGURE 3

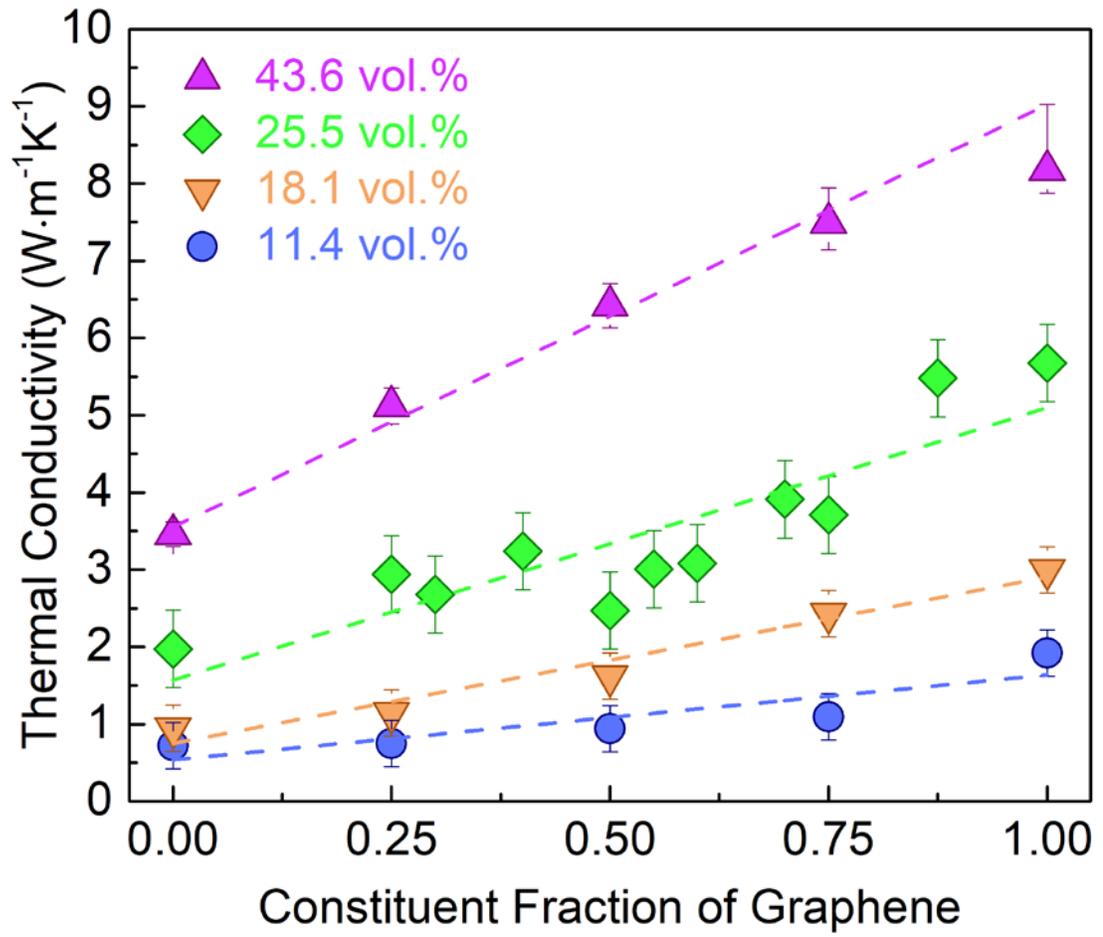

FIGURE 4

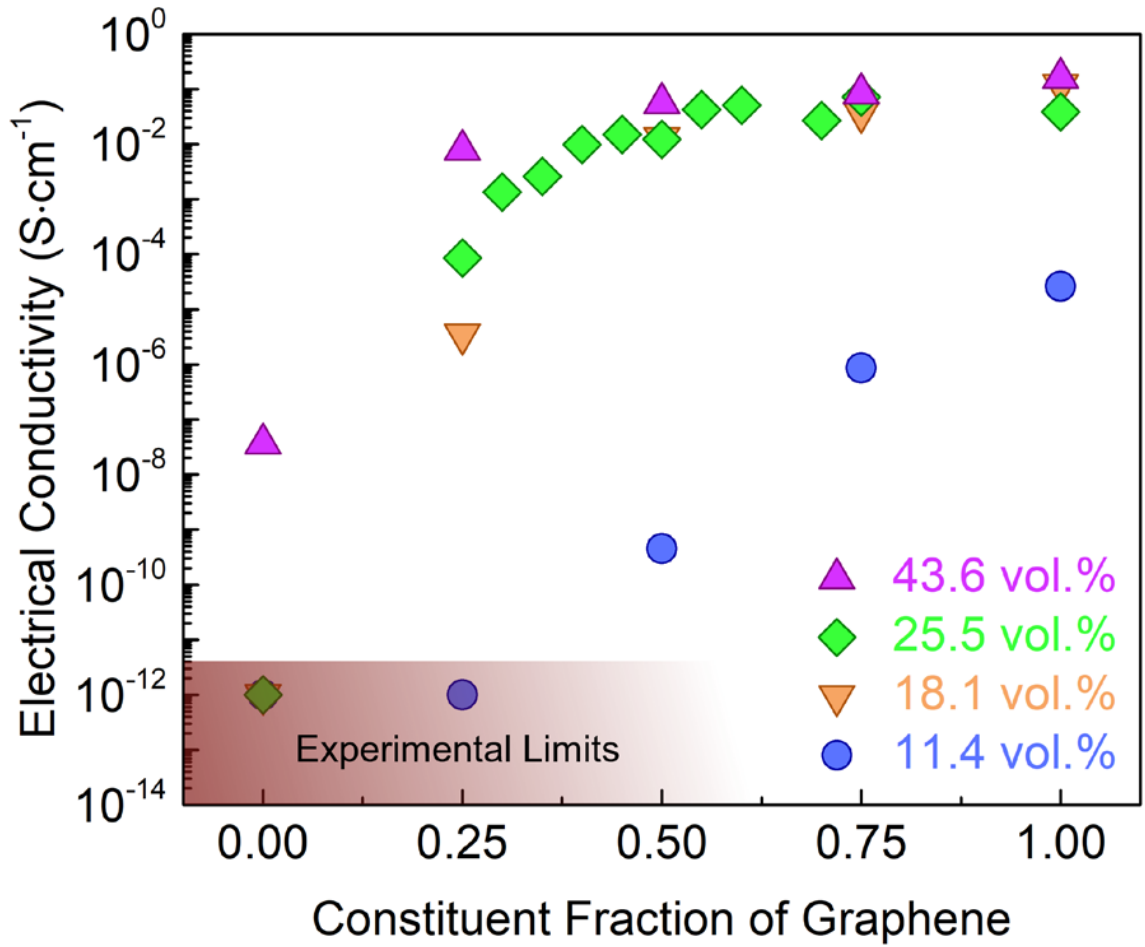

FIGURE 5